\documentclass[10pt]{iopart}

\usepackage{graphicx}
\usepackage{dcolumn}
\usepackage{bm}
\usepackage{color}

\begin{document}

\title[Non linear EIC]{Theoretical analysis of the saturation phase of the $1/1$ energetic-ion-driven resistive interchange mode}


\author{J. Varela}
\ead{jvrodrig@fis.uc3m.es}
\address{Universidad Carlos III de Madrid, 28911 Leganes, Madrid, Spain}
\address{National Institute for Fusion Science, National Institute of Natural Science, Toki, 509-5292, Japan}
\address{Oak Ridge National Laboratory, Oak Ridge, Tennessee 37831-8071, USA}
\author{D. A. Spong}
\address{Oak Ridge National Laboratory, Oak Ridge, Tennessee 37831-8071, USA}
\author{L. Garcia}
\address{Universidad Carlos III de Madrid, 28911 Leganes, Madrid, Spain}
\author{S. Ohdachi}
\address{National Institute for Fusion Science, National Institute of Natural Science, Toki, 509-5292, Japan}
\author{K. Y. Watanabe}
\address{National Institute for Fusion Science, National Institute of Natural Science, Toki, 509-5292, Japan}
\author{R. Seki}
\address{SOKENDAI, Department of Fusion Science, Toki/Gifu and National Institute for Fusion Science, National Institute of Natural Science, Toki, 509-5292, Japan}
\author{Y. Ghai}
\address{Oak Ridge National Laboratory, Oak Ridge, Tennessee 37831-8071, USA}

\date{\today}

\begin{abstract}
The aim of the present study is to analyze the saturation regime of the energetic-ion-driven resistive interchange mode (EIC) in the LHD plasma. A set of non linear simulations are performed by the FAR3d code that uses a reduced MHD model for the thermal plasma coupled with a gyrofluid model for the energetic particles (EP) species. The hellically trapped EP component is introduced through a modification of the averaged drift velocity operator to include their precessional drift. The non linear simulation results show similar $1/1$ EIC saturation phases with respect to the experimental observations, reproducing the enhancement of the $n/m = 1/1$ resistive interchange modes (RIC) amplitude and width as the EP $\beta$ increases, the EP $\beta$ threshold for the $1/1$ EIC excitation, the further destabilization of the $1/1$ EIC as the population of the helically trapped EP increases and the triggering of burst events. The frequency of the $1/1$ EIC calculated during the burst event is $9.4$ kHz and the $2/2$ and $3/3$ overtones are destabilized, consistent with the frequency range and the complex mode structure measured in the experiment. In addition, the simulation shows the inward propagation of the $1/1$ EIC due to the non linear destabilization of the $3/4$ and $2/3$ EPMs, leading to the partial overlapping between resonances during the burst event. Finally, the analysis of the $1/1$ EIC stabilization phase shows the excitation of the $1/1$ RIC as soon as the flattening induced by the $1/1$ EIC in the pressure profile vanishes, leading to the retrieval of the pressure gradient at the plasma periphery and the overcoming of the RIC stability limit.
\end{abstract}

%
%
%
%
%


\vspace{2pc}

\maketitle

\ioptwocol

\section{Introduction \label{sec:introduction}}

Low density and high ion temperature discharges in Large Helical Device (LHD) show magnetic fluctuations similar to the fishbone oscillations \cite{1,2}. These perturbations are observed if the plasma is strongly heated by perpendicular neutral beam injectors (NBI) \cite{3} and the resistive interchange modes (RIC) are unstable at the plasma periphery \cite{4,5,6}. These perturbations are called $1/1$ energetic-ion-driven resistive interchange mode (EIC), triggered by the resonance between the precession motion of the helically trapped EP generated by the perpendicular NBI (around the $10$ kHz) and the RIC \cite{7}.

The $1/1$ EIC can be included in the family of the energetic particle modes (EPM) \cite{8}, modes destabilized for frequencies in the shear Alfven continua if the continuum damping is not strong enough to stabilize them \cite{9,10}. The destabilization of the EPM should be avoided because the heating efficiency of the device decreases due to the loss of EP before thermalization.

Different phases are observed during the $1/1$ EIC saturation \cite{7}. First, the RIC is unstable before the excitation of the $1/1$ EIC once the density of helically trapped EP increases above a given threshold (EP $\beta$), identified in the experiment as a plasma perturbation with a frequency around $4$ kHz (Phase I). Next, the $1/1$ EIC is further destabilized as the EP $\beta$ increases leading to a larger mode width and the inward expansion of the perturbation (Phase II). Then, the frequency of the $1/1$ EIC suddenly increases to $9$ kHz and the instability shows a bursting behavior while a complex radial structure appears between $r/a = [0.6 , 0.85]$ as the perturbation keeps propagating inward (Phase III). Finally, a steady magnetic island appears at the radial location of the $1/1$ rational surface, the $1/1$ EIC stabilizes and the RIC is triggered again (Phase IV).

LHD is a helical device heated by three NBI lines parallel to the magnetic axis with an energy of $180$ keV and two NBI perpendicular to the magnetic axis with an energy of $32$ keV. Advanced LHD operation scenarios as high $\beta$ discharges require the intensive use of the perpendicular NBIs, leading to a large accumulation of helically trapped particles at the plasma periphery and the destabilization of $1/1$ EIC, limiting the maximum $\beta$ of the experiment.

The aim of the present study is to analyze the saturation of the $1/1$ EIC reproducing the different instability phases observed in the experiment, in particular the burst event. The simulations are performed using the FAR3D gyro-fluid code \cite{11,12,13,14}. The numerical model solves the reduced non linear resistive MHD equations coupled with the moment equations of the energetic ion density and parallel velocity \cite{15,16,17}. The linear wave-particle resonance effects required for Landau damping/growth are added for the appropriate Landau closure relations, as well as the parallel momentum response of the thermal plasma required for coupling to the geodesic acoustic waves \cite{18}. Six field variables evolves starting from an equilibria calculated by the VMEC code \cite{19}.

This paper is organized as follows. The model equations, numerical scheme and equilibrium properties are described in section \ref{sec:model}. The analysis of the non linear simulation during the saturation phase of the $1/1$ EIC is performed in section \ref{sec:sim}. Next, the conclusions of this paper are presented in section \ref{sec:conclusions}.

\section{Equations and numerical scheme \label{sec:model}}

Following the method employed in Ref.\cite{20}, a reduced set of equations for high-aspect ratio configurations and moderate $\beta$-values (of the order of the inverse aspect ratio) is derived retaining the toroidal angle variation based upon an exact three-dimensional equilibrium that assumes closed nested flux surfaces. The effect of the energetic particle population in the plasma stability is included through moments of the fast ion kinetic equation truncated with a closure relation \cite{21}, describing the evolution of the energetic particle density ($n_{f}$) and velocity moments parallel to the magnetic field lines ($v_{||f}$). The coefficients of the closure relation are selected to match analytic TAE growth rates based upon a two-pole approximation of the plasma dispersion function. A detail description of the model equations and numerical scheme is included in the appendix.

The present model was already used to study the linear stability of LHD plasma, particularly the AE destabilization by EP \cite{22}, AE stability of plasma with multiple EP populations \cite{23}, the effect of the tangential NBI current drive on the stability of pressure and EP driven MHD modes \cite{24} as well as EIC stabilization strategies \cite{25}, showing a reasonable agreement with the observations. In addition, non linear studies were performed reproducing the sawtooth-like and internal collapse events observed in LHD plasma \cite{26,27,28,29}.

\subsection{Equilibrium properties}

A fixed boundary equilibrium from the VMEC code \cite{19} was calculated during the LHD shot $116190$ after the destabilization of an EIC event including the EP component of the pressure. The electron density and temperature profiles were reconstructed by Thomson scattering data and electron cyclotron emission. Table~\ref{Table:1} shows the main parameters of the thermal plasma. The cyclotron frequency is $\omega_{cy} = 2.41 \cdot 10^{8}$ s$^{-1}$.

\begin{table}[t]
\centering
\begin{tabular}{c c c c}
\hline
$T_{i}$ (keV) & $n_{i}$ ($10^{20}$ m$^{-3}$) & $\beta_{th}$ ($\%$) & $V_{A}$ ($10^{7}$ m/s) \\ \hline
2 & 0.25 & 0.32 & 1.1 \\
\end{tabular}
\caption{Thermal plasma properties in the reference model (values at the magnetic axis). The first column is the thermal ion temperature, the second column is the thermal ion density, the third column is the thermal $\beta$ and the fourth column is the Alfv\' en velocity.} \label{Table:1}
\end{table}

\begin{figure}[h!]
\centering
\includegraphics[width=0.45\textwidth]{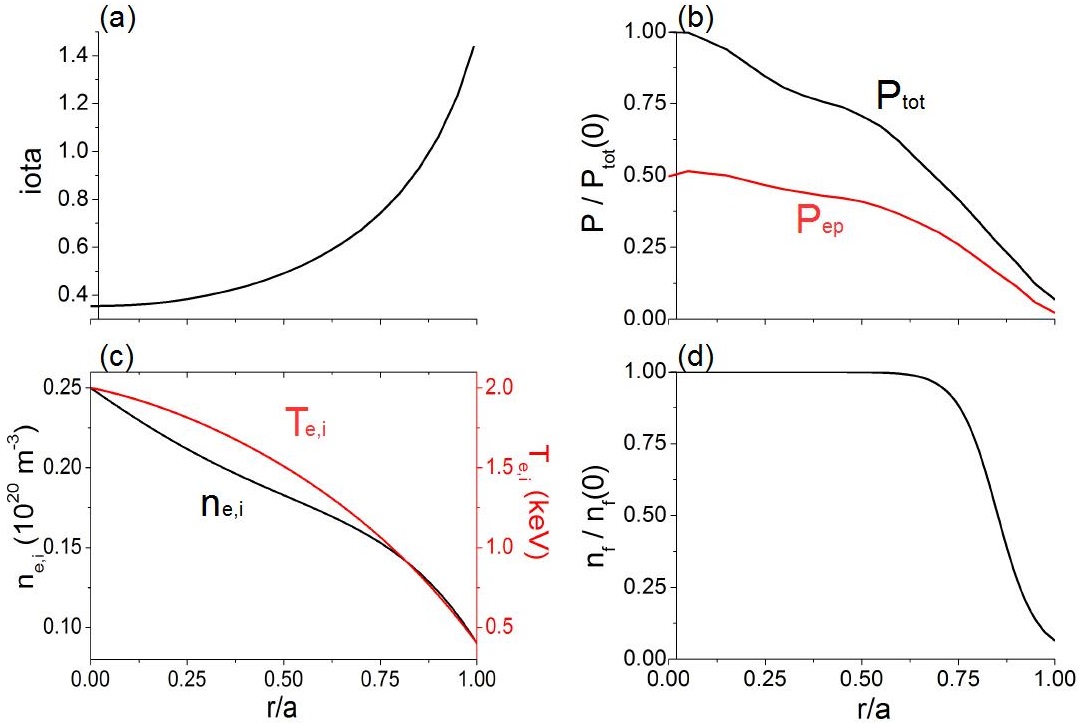}
\caption{(a) Iota profile, (b) normalized pressure (thermal + EP pressure) profile, (c)  thermal plasma density and temperature profiles and (d) density profile of the helically trapped EP.}\label{FIG:1}
\end{figure}

The magnetic field at the magnetic axis is $2.5$ T and the averaged inverse aspect ratio is $\varepsilon=0.16$. The energy of the injected particles by the perpendicular NBI is $T_{f,\perp}(0) = 40$ keV, but we take the nominal energy $T_{f,\perp}(0) = 28$ keV ($v_{th,f} = 1.64 \cdot 10^{6}$ m/s) resulting in an averaged Maxwellian energy equal to the average energy of a slowing-down distribution. Figure~\ref{FIG:1} (a) shows the iota profile, (b) the normalized pressure profile, (c) the thermal plasma density (black line) and temperature (red line) profiles and (d) the EP density profile. It should be noted that the effect of the equilibrium toroidal rotation is not included in the model for simplicity. The effect of the Doppler shift on the instability frequency caused by the toroidal rotation is small, particularly for a mode located in the plasma periphery, as it is observed in the experiments.

\subsection{Simulations parameters}

The dynamic and equilibrium toroidal ($n$) and poloidal ($m$) modes included in the study are summarized in table~\ref{Table:2}. The mode selection covers the resonant rational surfaces between the middle plasma and the plasma periphery, thus the inner plasma region ($r/a < 0.5$) is not included in the simulations in order to reduce the computational time and avoid the destabilization of undesired modes in the plasma core. The simulations are performed with a uniform radial grid of 400 points. In the following, the mode number notation is $n/m$, which is consistent with the $\iota=n/m$ definition for the associated resonance.

\begin{table}[h]
\centering
\begin{tabular}{c | c }
\hline
$n$ & $m$  \\ \hline
$0$ & $[0 , 6]$ \\
$1$ & $[1 , 2]$ \\
$2$ & $[2 , 4]$ \\
$3$ & $[2 , 6]$ \\
\end{tabular}
\caption{Dynamic and equilibrium toroidal (n) and poloidal (m) modes in the simulations.} \label{Table:2}
\end{table}

The closure of the kinetic moment equations (6) and (7) breaks the MHD parities so both parities must be included for all the dynamic variables. The convention of the code with respect to the Fourier decomposition is, in the case of the pressure eigenfunction, that $n > 0$ corresponds to $cos(m\theta + n\zeta)$ and $n<0$ corresponds to $sin(-m\theta - n\zeta)$. For example, the Fourier component for mode $3/4$ is $\cos(3\theta + 4\zeta)$ and for the mode $-3/-4$ is $\sin(3\theta + 4\zeta)$. The magnetic Lundquist number is assumed $S=5\cdot 10^6$, consistent with the S value at the plasma periphery of LHD plasma, reproducing the stability properties regarding the plasma resistivity for the 1/1 resistive interchange mode (RIC) and $1/1$ EIC.

The density and temperature of the population of helically trapped EP generated by the perpendicular beam are calculated by the code MORH \cite{30,31}. For simplicity, no radial dependency of the EP energy is considered and the EP density profile given by the MORH code is fitted to the following analytic expression:

\begin{equation}
\label{EP_dens}
$$n_{f,\perp}(r) = \frac{(0.5 (1+ \tanh(\delta_{r} \cdot (r_{peak}-r))+0.02)}{(0.5 (1+\tanh(\delta_{r} \cdot r_{peak}))+0.02)}$$
\end{equation}
with the location of the EP density gradient profile defined by the variable $r_{peak} = 0.85$ and the flatness by $\delta_{r} = 10$. A previous study indicated the passing particles generated by the tangential NBIs could have a stabilizing effect of the EIC \cite{25}. Nevertheless, the EIC saturation phase can be studied independently of the passing particle effects, reason why the passing particles are not included in the simulations for simplicity.

The Landau closure in the model is based on two moment equations for the energetic particles, which is equivalent to a two-pole approximation of the plasma dispersion relation. The closure coefficients are adjusted by fitting analytic AE growth rates. Such assumptions are consistent with a Lorentzian energy distribution function for the energetic particles. The lowest order Lorentzian can be matched either to a Maxwellian or to a slowing-down distribution by choosing an equivalent average energy. For the results given in this paper, we have matched the EP temperature to the mean energy of a slowing-down distribution function.

\begin{figure}[h!]
\centering
\includegraphics[width=0.45\textwidth]{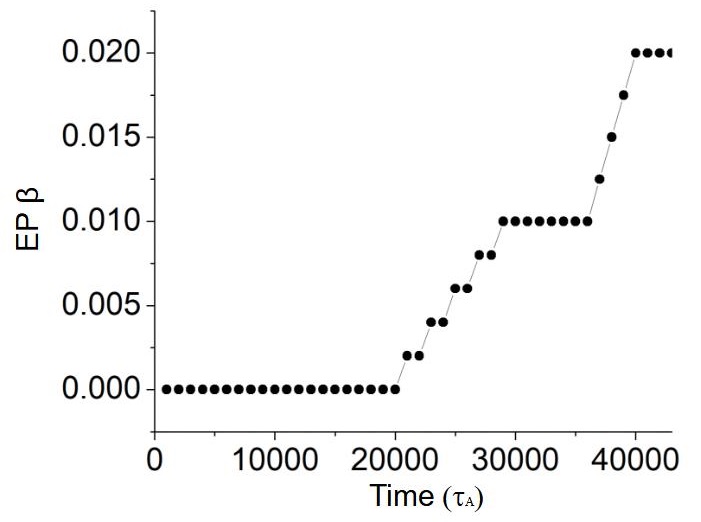}
\caption{EP $\beta$ along the simulation.}\label{FIG:2}
\end{figure}

The simulations begin with a thermal $\beta$ $50 \%$ smaller with respect to the original equilibria, increasing to $60 \%$ after $t = 10^{4} \tau_{A0}$. The RIC is unstable for $60 \%$ of the original thermal $\beta$ thus the value of the thermal $\beta$ is no further increased during the simulation, avoiding numerical instabilities and the destabilization of the $1/2$ interchange mode in the middle plasma region. Once the non linear simulation for the thermal plasma reaches $t = 2 \cdot 10^{4} \tau_{A0}$, the EP destabilizing effect is turned on by linearly increasing the EP $\beta$. Figure~\ref{FIG:2} indicates the EP $\beta$ values along the simulation. The EP $\beta$ in the experiment increases along the discharge because the perpendicular NBI generates a larger population of trapped EP, thus the EP $\beta$ may increase along the simulation to reproduce the different phases of the $1/1$ EIC observed in the experiment, linked to different EP $\beta$ thresholds. The EP $\beta$ increments in the simulation are selected to avoid large variations of the AE/EPM stability in short simulation times, that is to say, the simulation can capture the effect of the EP $\beta$ on the AE/EPM stability, identifying the EP $\beta$ threshold linked to the different phases of the EIC. Between $t = 2.0 - 2.8 \cdot 10^{4} \tau_{A0}$ the EP $\beta$ increases from $0.0$ to $0.01$ by $\Delta \beta = 0.002$ each $2000 \tau_{A0}$, reproducing the Phase I and the early phase II of the $1/1$ EIC. Next, the EP $\beta$ is fixed to $0.01$ until $t = 3.6 \cdot 10^{4} \tau_{A0}$, reproducing the phase II. Finally,the EP $\beta$ increases by $\Delta \beta = 0.0025$ between $t = 3.7 - 4.0  \cdot 10^{4} \tau_{A0}$ up to $0.02$ leading to the triggering of a burst event (Phase III). The stabilization phase of the $1/1$ EIC is analyzed separately (Phase IV), performing by a new simulation including only the $n=1$ toroidal family for simplicity. 

\section{Saturation of the $1/1$ EIC \label{sec:sim}}

\begin{figure}[h!]
\centering
\includegraphics[width=0.45\textwidth]{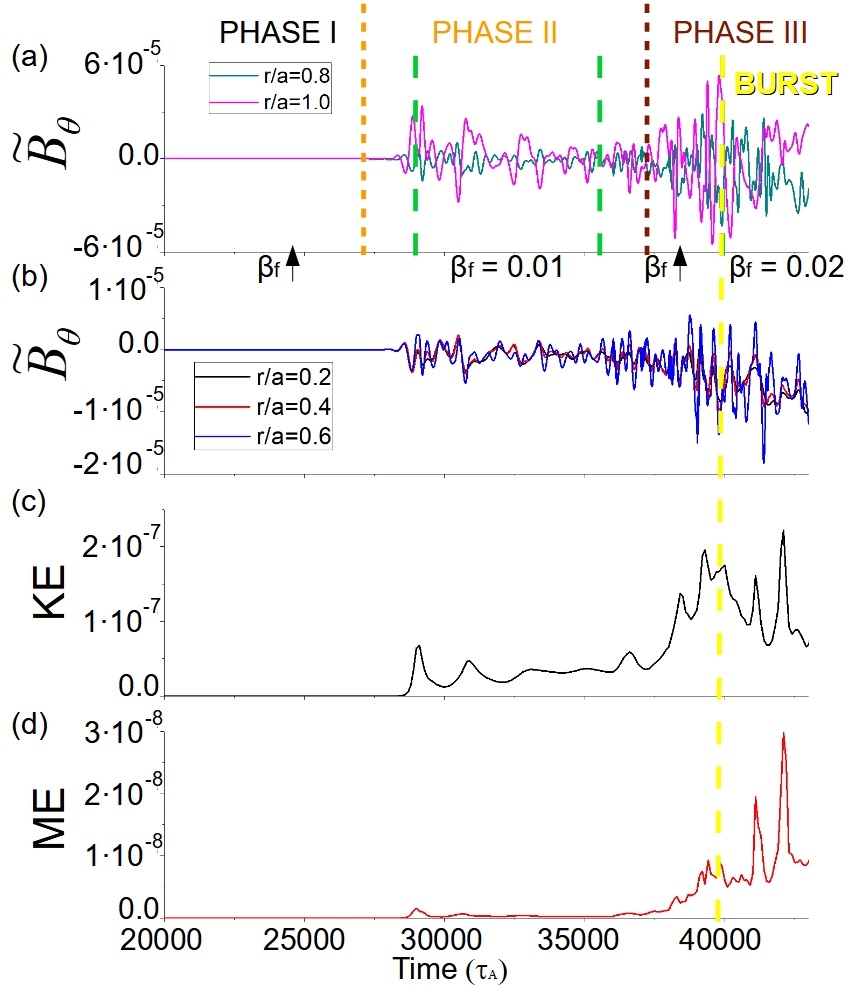}
\caption{Time evolution of the poloidal component of the magnetic field perturbation (a) at $r/a = 1.0$ (pink line) and $0.8$ (cyan line), (b) inner-middle plasma at $r/a = 0.6$ (blue line), $0.4$ (red line) and $0.2$ (black line), (c) normalized kinetic energy and (d) normalized magnetic energy. The dashed orange line indicates the beginning of the EIC Phase II and the dashed dark red line the beginning of the Phase III. The long dashed green line shows the simulation time when the EP $\beta$ linearly increases. The yellow long dashed line indicates the simulation time when the burst even is triggered.}\label{FIG:3}
\end{figure}

The $1/1$ EIC phases I to III are reproduced by a single non linear simulation as the EP $\beta$ increases. Figure~\ref{FIG:3}, panels a and b, show the time evolution of the poloidal component of the magnetic field perturbation ($\tilde B_{\theta}$) at different plasma regions. A fast oscillating perturbation appears between $r/a = 0.8 - 1.0$ (pink and cyan lines in panel a), indicating the destabilization of the $1/1$ EIC in the plasma periphery at $t = 2.6 \cdot 10^{4} \tau_{A0}$ and the transition from the Phase I to the Phase II . The perturbation amplitude increases during the Phase II and reaches its maximum amplitude during the burst event triggered in the Phase III ($t = 3.9 \cdot 10^{4} \tau_{A0}$). In addition, the perturbation extends from the plasma periphery to the middle plasma region (blue line in panel b). Regarding the kinetic and magnetic energy (KE and ME, respectively) of the modes (including the $n=0$ toroidal family contribution), there are two local maximum observed after the $1/1$ EIC is excited and during the burst event, although the KE/ME energy peak is $3$ times larger in the burst event.

\begin{figure}[h!]
\centering
\includegraphics[width=0.45\textwidth]{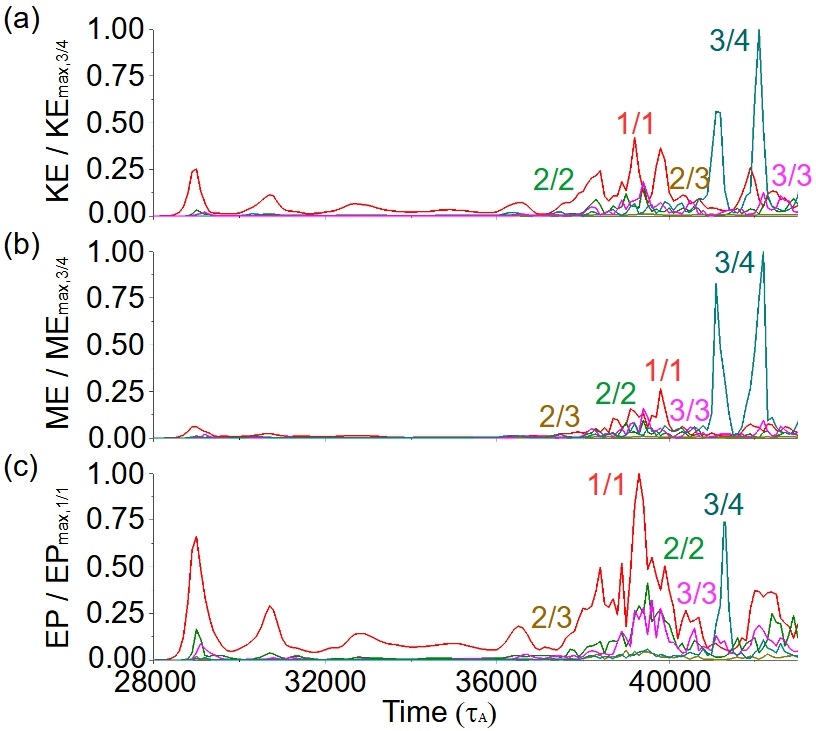}
\caption{Time evolution of the dominant modes normalized (a) kinetic (b) magnetic and (c) EP energy. The mode $1/1$ is indicated by the red line, the $2/2$ by the green line, the $3/3$ by the pink line, the $2/3$ by the brown line and the $3/4$ by the cyan line.}\label{FIG:4}
\end{figure}

Figure~\ref{FIG:4} indicates the time evolution of the normalized energy for the dominant modes among the phases II and III of the simulation, excluding the mode $0/0$. The consecutive local maxima / minima of the $1/1$ mode energy (red line) corresponds to cycles of further destabilization / saturation of the $1/1$ EIC, correlated with an enhancement / weakening of $\tilde B_{\theta}$ oscillations at the plasma periphery. It should be noted that the local maxima of the $1/1$ mode are linked to local maxima of the overtones $2/2$ (green line) and $3/3$ (pink line) energy, particularly large during the bursting event. From $t = 3 \cdot 10^{4} \tau_{A0}$ there are local maxima of the $3/4$ mode energy indicating the mode destabilization. Once the $3/4$ mode is unstable, $3/4$ mode kinetic (panel a) and magnetic (panel b) energy show a large enhancement at the end of the phase II, larger regarding $1/1$ mode kinetic and magnetic energy during the phase III. Consequently, the largest perturbation of the surrounding plasma flows as well as the strongest magnetic perturbation is caused by the $3/4$ mode during the phase III, although the dominant perturbations during the phases I and II are generated by the mode $1/1$. Likewise, the energy associated to the EP destabilizing effect (panel c) is largely dominated by the $1/1$ mode and overtones during the phases I and II, indicating that there is an energy transfer from the $1/1$ EIC towards the $1/1$ mode overtones as well as the $3/4$ mode. On the other hand, the leading destabilizing effect is caused by the $3/4$ mode after the burst event.

In the following, each $1/1$ EIC phase is analyzed separately.

\subsection{Phase I: $1/1$ EIC destabilization}

\begin{figure}[h!]
\centering
\includegraphics[width=0.5\textwidth]{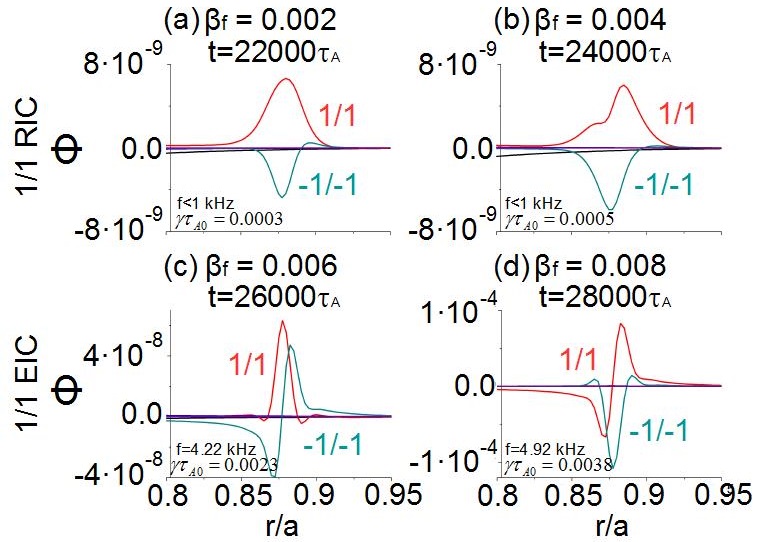}
\caption{$1/1$ mode electrostatic potential eigenfunction calculated in the non linear simulation if (a) EP $\beta = 0.002$, (b) $0.004$, (c) $0.006$ and (d) $0.008$. The mode growth rate and frequency is included.}\label{FIG:5}
\end{figure}

Figure~\ref{FIG:5} indicates the evolution of the electrostatic potential eigenfunction of the $1/1$ mode during the Phase I and early Phase II of the non linear simulation. In addtion, the growth rate and frequency is calculated performing linear simulations using the profiles of the non linear simulation. If the EP $\beta \le 0.004$ the mode is a RIC (panels a and b) because the eigenfunction is symmetric with respect to the mode parities and the frequency is lower than $1$ kHz. The RIC is triggered if the pressure profile gradient is large enough to exceed the stability limit of the interchange modes at the plasma periphery, although the RIC amplitude, width and growth rate is also affected by the driving of the EP, increasing as the EP $\beta$ grows. On the other hand, if the EP $\beta = 0.006$ the eigenfunction of the $1/1$ mode parities are not symmetric, the mode frequency is $4$ kHz and the growth rate increases almost one order of magnitude, indicating the destabilization of the $1/1$ EIC. The $1/1$ EIC amplitude, width and growth rate also increases with the EP $\beta$ although this dependency is stronger regarding the RIC.

It should be noted that the EP $\beta$ threshold to destabilize the $1/1$ EIC is $0.006$ in the simulation, similar to the critical $\beta$ measured in the experiment. In addition, the increase of the amplitude and width of the RIC and $1/1$ EIC is also observed in the experiment as the EP population grows.

\subsection{Phase II: inward propagation of the instability}

\begin{figure}[h!]
\centering
\includegraphics[width=0.5\textwidth]{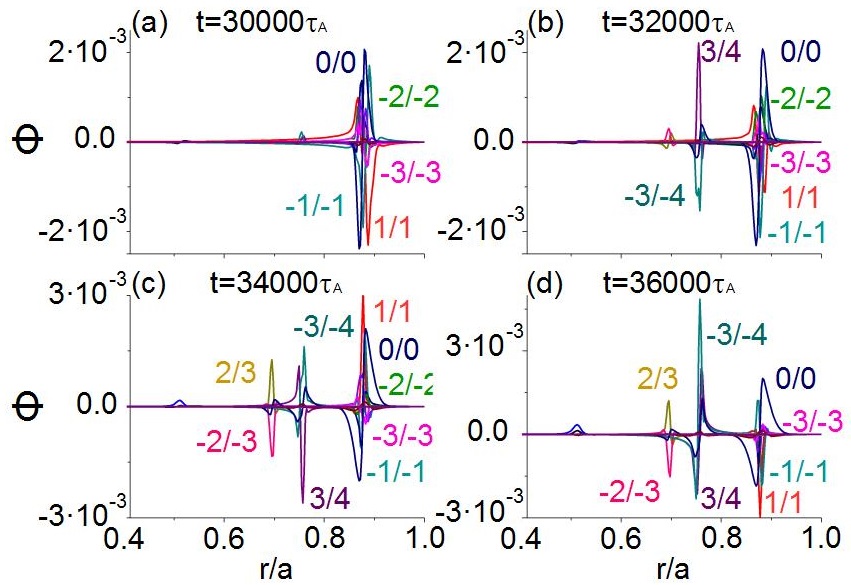}
\caption{Electrostatic potential eigenfunction of the $n=1$ to $n=3$ toroidal mode families calculated in the non linear simulation at (a) $t = 3.0 \cdot 10^{4} \tau_{A}$, (b) $t =3.2 \cdot 10^{4} \tau_{A}$, (c) $t =3.4 \cdot 10^{4} \tau_{A}$ and (d) $t =3.6 \cdot 10^{4} \tau_{A}$.}
\label{FIG:6}
\end{figure}

The saturation of the $1/1$ EIC begins in the Phase II as soon as the non linear effects dominate the instability evolution, as can be observed in the eigenfunction of the electrostatic potential of the $n=0$ to $3$ toroidal mode families (fig.~\ref{FIG:6}). At $t = 3.0 \cdot 10^{4} \tau_{A}$, panel a, the $1/1$ EIC amplitude increases and the overtones $2/2$ and $3/3$ are unstable. In addition, the mode $0/0$ is destabilized indicating that the $1/1$ EIC modifies the nearby plasma flows generating the so-called shear flows, that is to say, part of the $1/1$ EIC energy is transferred toward the thermal plasma generating a readjustment of the equilibrium flows in the plasma periphery. A detailed analysis of the generation of shear flows by the $1/1$ EIC is out of the scope of this article and will be the topic of a future study. At present, the analysis of the simulation data presumes the stability of the $1/1$ EIC is affected by the presence of shear flows, particularly during the saturation phase of the instability once a feedback effect between the mode and the shear flows exit. At $t = 3.2 \cdot 10^{4} \tau_{A}$, panel b, the mode $3/4$ is triggered at $r/a = 0.76$ due to the non-linear coupling with the mode $1/1$, so that a fraction of the $1/1$ EIC energy is also channeled toward modes of different toroidal families. The same way, at $t = 3.4 \cdot 10^{4} \tau_{A}$, panel c, the mode $2/3$ is triggered around $r/a = 0.7$. Later, at $t = 3.6 \cdot 10^{4} \tau_{A}$, panel d, the amplitude of the $3/4$ mode keeps increasing indicating that the energy channeling continues while the $1/1$ EIC is active. Consequently, the enhancement of the electron temperature perturbation measured at different radial location during the Phase II \cite{7} could be caused by the combined effect of the further destabilization of the $1/1$ EIC (increase of the $1/1$ mode width), the excitation of the $2/2$ and $3/3$ overtones as well as the non linear destabilization of the $3/4$ and $2/3$ modes, leading to the inward propagation of the perturbation.

Table~\ref{Table:3} shows the growth rate and frequency of the $n=1$ to $3$ dominant modes (the modes with the largest growth rate of each toroidal family) during the Phase II, calculated by linear simulations using the profiles obtained from the non linear simulations. The mode $3/4$ has a frequency around $12$ kHz and the eigenfunction of the mode parities are asymmetric, indicating that the $3/4$ mode is an EPM. On the other hand, the mode $2/3$ has a frequency below $1$ kHz, a symmetric eigenfuncion with respect to mode parities and a growth rate $5$ times smaller than the $1/1$ EIC, thus the $2/3$ is an interchange mode.
\begin{table}[h]
\centering
\begin{tabular}{c | c c c}
\hline
$n/m$ & Time & Growth rate & f  \\ 
      & ($10^{4}\tau_{A0}$) & ($\gamma \tau_{A0}$) & (kHz) \\ \hline
1/1 & $2.9$ & $0.0067$ & $7.16$ \\
1/1 & $3.0$ & $0.0057$ & $6.15$ \\
1/1 & $3.2$ & $0.0059$ & $6.39$ \\
3/4 & $3.2$ & $0.0079$ & $12.21$ \\
1/1 & $3.4$ & $0.0058$ & $6.32$ \\
2/3 & $3.4$ & $0.0010$ & $< 1$ \\
3/4 & $3.4$ & $0.0078$ & $12.23$ \\
1/1 & $3.6$ & $0.0058$ & $6.26$ \\
2/3 & $3.6$ & $0.0010$ & $< 1$ \\
3/4 & $3.6$ & $0.0082$ & $12.59$ \\
\end{tabular}
\caption{Growth rate and frequency of the $n=1$ to $n=3$ modes during the Phase II.} \label{Table:3}
\end{table}

\begin{figure}[h!]
\centering
\includegraphics[width=0.45\textwidth]{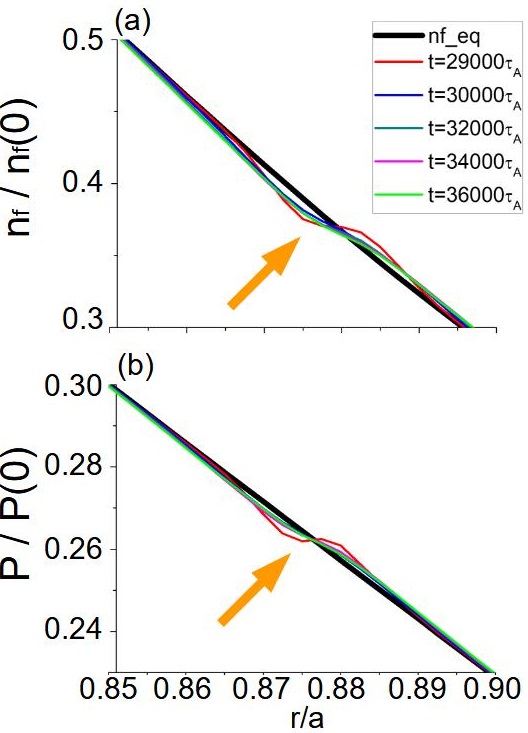}
\caption{Evolution of the (a) EP density and (b) pressure profiles towards the Phase II (Equilibrium + $0/0$ perturbation component). The black broad line indicates the equilibrium profiles. The orange arrow indicates the radial location of the largest deviation from the equilibrium.}
\label{FIG:7}
\end{figure}

Now, the effect of the $1/1$ EIC on the equilibrium profiles is analyzed. Figure~\ref{FIG:7} shows the evolution of the EP density and pressure profiles during the Phase II. There is a flattening of the EP density and pressure profiles around $r/a = 0.88$, the radial location of the $1/1$ rational surface. The largest deviation from the equilibrium profiles is observed at $t = 2.9 \cdot 10^{4}\tau_{A0}$ once the $1/1$ EIC growth rate reaches a local maximum (also identified by the local maxima of the KE and ME, see fig~\ref{FIG:4}). The $1/1$ EIC saturates because the gradient of the EP density profile around the $1/1$ rational surface is weaker, leading to a decrease of the free energy required to sustain the $1/1$ EIC unstable. For this reason the mode growth rate decreases. It should be noted that the EP $\beta$ of the simulation is not modified after the $1/1$ EIC is triggered, even though the experimental observations indicate that part of the helically trapped EP population is lost after the plasma relaxation. The EP $\beta$ of the simulation does not decrease because the destabilizing effect of the EP is maximized to reduce the simulation time required to reproduce the different $1/1$ EIC phases. With regard to the effect of the $1/1$ EIC on the pressure, the pressure is also flattened around the $1/1$ rational surface, leading to a decrease of the pressure gradient at the plasma periphery. Thus, the stability of the EPM/AE and pressure gradient driven modes are non linearly connected, in a manner resulting in the stabilization of the RIC during the saturation of the $1/1$ EIC. The displacement of $\Delta (r/a) = 0.0025$ between the flattening induced in the pressure and EP density profiles is caused by the model resolution.

\begin{figure}[h!]
\centering
\includegraphics[width=0.45\textwidth]{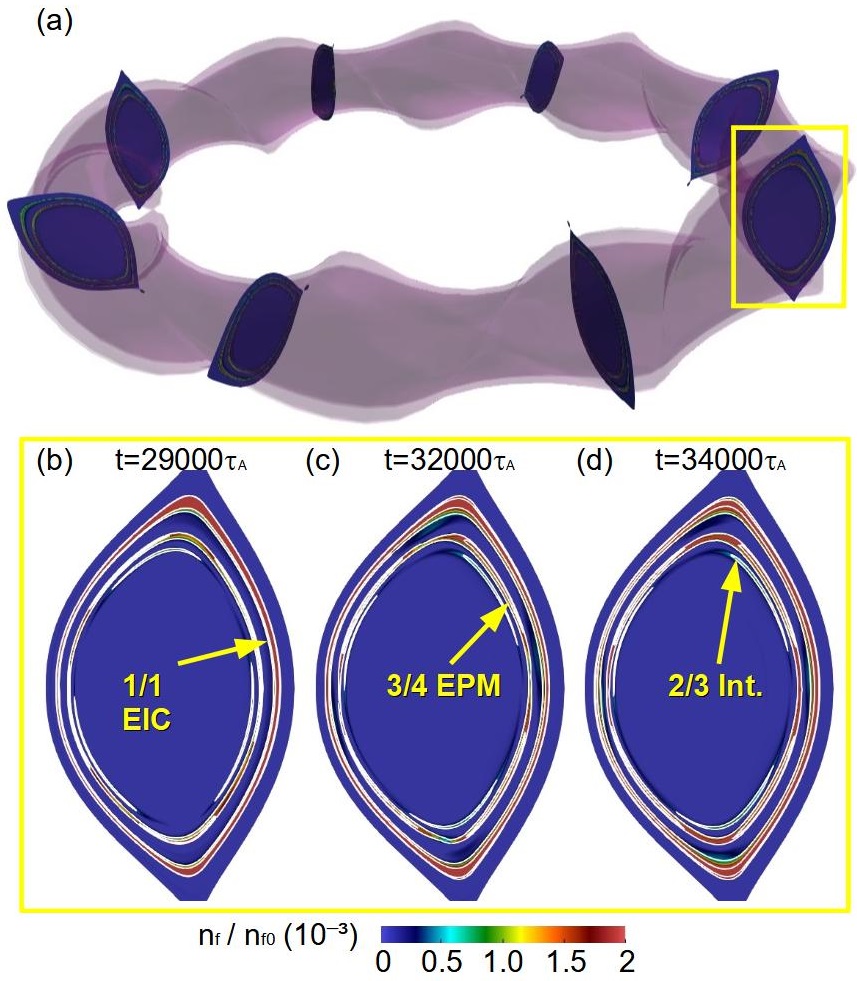}
\caption{(a) Isosurface and poloidal contours of the EP density. The yellow rectangle indicates the location of the poloidal contour plotted. Evolution of the poloidal contour of the EP density perturbation at (b) $t =2.9 \cdot 10^{4} \tau_{A}$, (c) $t =3.2 \cdot 10^{4} \tau_{A}$ and (d) $t =3.4 \cdot 10^{4} \tau_{A}$. The yellow arrows indicate the location of the instabilities.}
\label{FIG:8}
\end{figure}

Figure~\ref{FIG:8} shows an isosurface and the evolution of the poloidal contour of the EP density perturbation. The $1/1$ EIC perturbation is located at the plasma periphery (panel b). There are two other perturbations linked to the $3/4$ EPM ($t =3.2 \cdot 10^{4} \tau_{A}$, panel c) and the $2/3$ interchange mode ($t =3.4 \cdot 10^{4} \tau_{A}$, panel d), although the strongest perturbation is caused by the $1/1$ EIC along the Phase II. 

In summary, the inward propagation of the instability observed in the experiment during the Phase II of the EIC saturation could be caused by the non linear destabilization of modes from different toroidal families, particularly the $3/4$ EPM and the $2/3$ interchange mode. These modes are excited due to the energy transferred from the $1/1$ EIC, channeled toward the thermal plasma for the case of the $2/3$ interchange mode and through non linear couplings with the $3/4$ EPM.

\subsection{Phase III: burst event}

\begin{figure}[h!]
\centering
\includegraphics[width=0.5\textwidth]{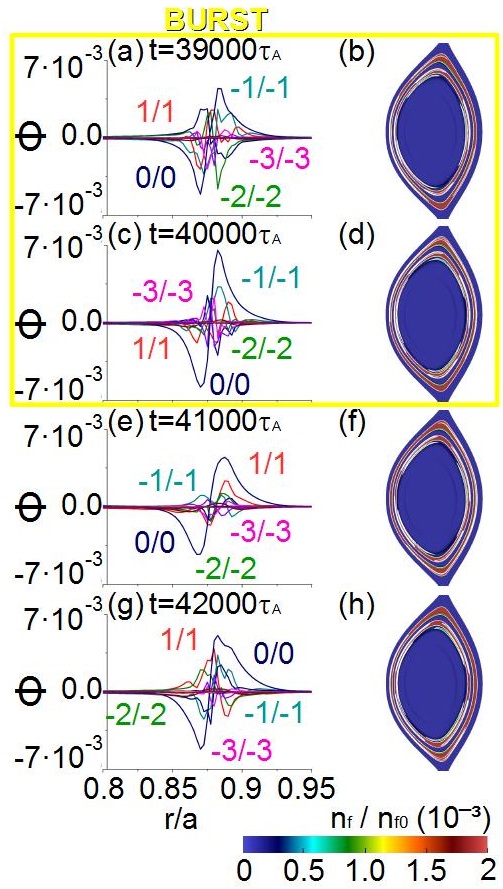}
\caption{Evolution of the electrostatic potential eigenfunction and poloidal contour of the EP density perturbation in the non linear simulation at (a) $t =3.8 \cdot 10^{4} \tau_{A}$, (c) $t = 3.9 \cdot 10^{4} \tau_{A}$, (e) $t =4.0 \cdot 10^{4} \tau_{A}$, (g) $t =4.1 \cdot 10^{4} \tau_{A}$ and (i) $t =4.2 \cdot 10^{4} \tau_{A}$. The yellow box indicates the modes eigenfunction and the poloidal contour of the EP density perturbation during the bursting event.}
\label{FIG:9}
\end{figure}

The burst event during the Phase III is triggered once the EP $\beta$ of the simulation increases to $0.02$, between $t =3.9-4.0 \cdot 10^{4} \tau_{A}$ (see fig~\ref{FIG:3} and~\ref{FIG:4}). The $2/2$ and $3/3$ overtones are further destabilized showing a mode amplitude similar to the mode $1/1$ during the burst event (fig~\ref{FIG:9}, panels a, c, e and g). In addition, the amplitude of the $0/0$ mode is almost $2$ times larger with respect to the $1/1$ mode ($t =4.0 \cdot 10^{4} \tau_{A}$), pointing out a strong perturbation of the equilibrium and the triggering of a large plasma relaxation, consistent with the local maxima of the KE and ME (see fig~\ref{FIG:3}c and d). The energy channeled from the $1/1$ EIC toward the thermal plasma and by non linear coupling with other toroidal families leads to the further destabilization of the middle-outer plasma region (please compare the poloidal contours of the EP density perturbation between the panels b, d, f and h). Table~\ref{Table:4} shows the growth rate and frequency of the $n=1$ to $3$ dominant modes during the burst event. At $t =3.9 \cdot 10^{4} \tau_{A}$ the growth rate of the mode $1/1$ is two times larger with respect to the Phase II and the frequency increases to $9.36$ kHz. In addition, the dominant modes of the $n=2$ and $3$ toroidal families are the overtones $2/2$ and $3/3$ with frequencies of $12.8$ and $22$ kHz, respectively. The overtones frequency and the asymmetry of the eigenfunction of the mode parities indicate that these modes are EPMs. Consequently, the simulation reproduces the increase of the $1/1$ EIC frequency to $9.4$ kHz measured in the experiment, as well as the instability enhancement at the plasma periphery ($0.88 < r/a < 0.91$) and the complex mode structure between $0.6 < r/a < 0.85$ during the burst event. It should be noted that at $t =4.0 \cdot 10^{4} \tau_{A}$ the dominant mode of the $n=3$ toroidal family is the $3/4$, consistent with the inward displacement of the instability.

\begin{table}[h]
\centering
\begin{tabular}{c | c c c}
\hline
$n/m$ & Time & Growth rate & f  \\ 
      & ($10^{4}\tau_{A0}$) & ($\gamma \tau_{A0}$) & (kHz) \\ \hline
1/1 & $3.9$ & $0.0119$ & $9.36$ \\
2/2 & $3.9$ & $0.0137$ & $12.82$ \\
3/3 & $3.9$ & $0.0221$ & $22.06$ \\
1/1 & $4.0$ & $0.0135$ & $9.53$ \\
2/2 & $4.0$ & $0.0170$ & $13.30$ \\
3/4 & $4.0$ & $0.0267$ & $18.15$ \\
\end{tabular}
\caption{Growth rate and frequency of the $n=1$ to $n=3$ dominant modes during the burst event.} \label{Table:4}
\end{table}

\begin{figure}[h!]
\centering
\includegraphics[width=0.45\textwidth]{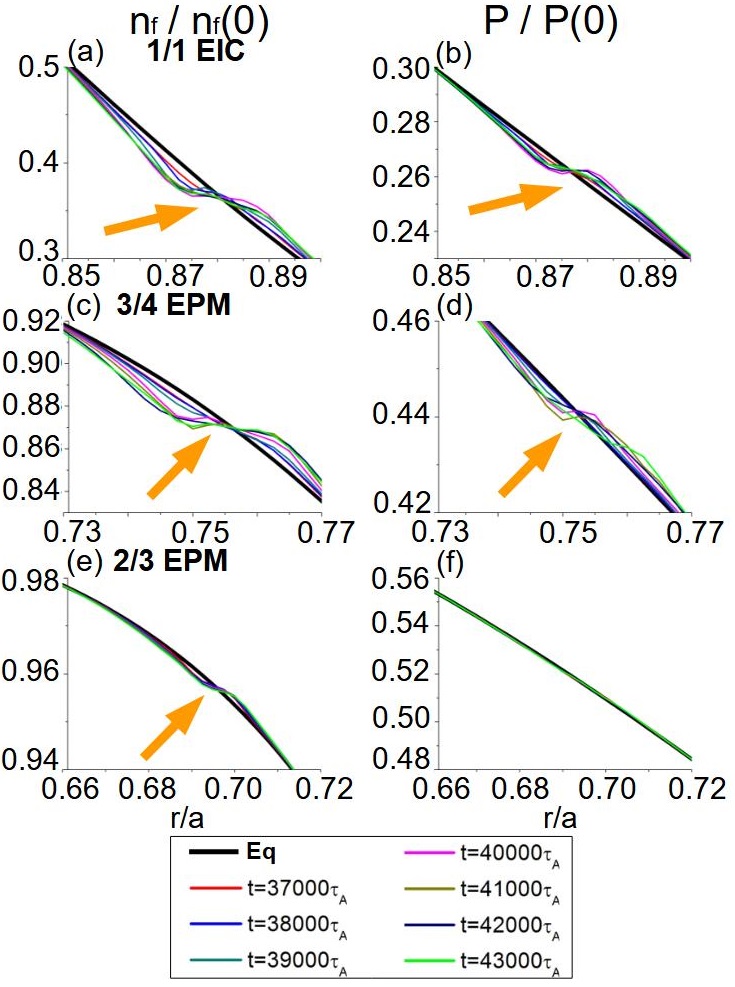}
\caption{Evolution of the (a) EP density and (b) pressure profiles of the $1/1$ EIC, Evolution of the (c) EP density and (d) pressure profiles of the $3/4$ EPM. Evolution of the (e) EP density and (f) pressure profiles of the $2/3$ EPM (Equilibrium + $0/0$ perturbation component). The black broad line indicates the equilibrium profiles. The orange arrow indicates the radial location of the strongest deviation from the equilibrium.}\label{FIG:10}
\end{figure}

\begin{figure}[h!]
\centering
\includegraphics[width=0.45\textwidth]{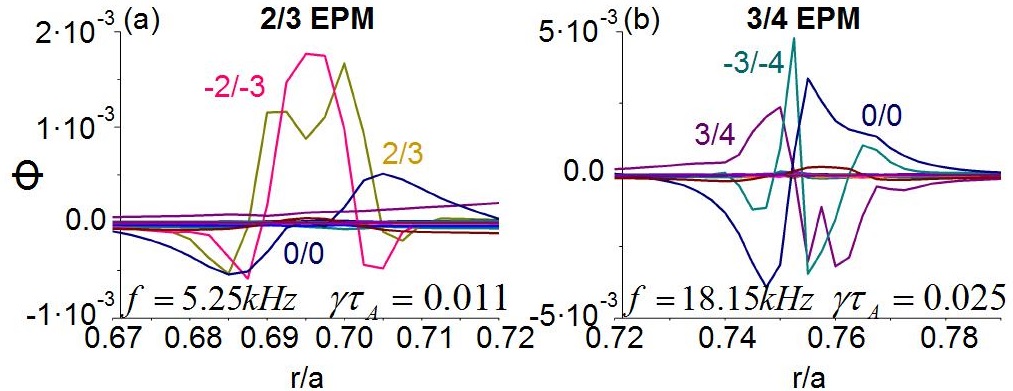}
\caption{Electrostatic potential eigenfunction in the non linear simulation of the (a) $2/3$ EPM and (b) $3/4$ EPM at $t = 4.0 \cdot 10^{4} \tau_{A}$.}
\label{FIG:11}
\end{figure}

Figure~\ref{FIG:10} shows the evolution of the EP density and pressure profiles in the Phase III. The flattening of the profiles around the $1/1$ rational surface is broader with respect to the Phase II (panels a and b), particularly once the burst event is triggered at $t = 3.9 \cdot 10^{4} \tau_{A}$. There is further flattening in the EP density and pressure profiles caused by the rational surface $3/4$ around $r/a = 0.76$ (panels c and d), showing the largest exclusion with respect to the equilibrium profiles at $t = 4.0 \cdot 10^{4} \tau_{A}$. There is a third flattening induced by the $2/3$ rational surface around $r/a = 0.69$, showing a smaller deviation from the equilibrium profiles with respect to the $1/1$ and $3/4$ rational surfaces, mainly observed in the EP density profile, indicating that the $2/3$ interchange mode destabilized during the phase II evolved to a $2/3$ EPM due to a the non linear coupling with the $1/1$ EIC. Figure~\ref{FIG:11} shows the electrostatic potential eigenfunction of the $2/3$ and $3/4$ EPMs, growth rate and frequency. The destabilization of the mode $0/0$ by the $2/3$ and $3/4$ EPMs indicates that these modes also generate a local modification of the surrounding plasma flows, although the effect is weaker than for the $1/1$ EIC because the $0/0$ mode amplitude is smaller. Consequently, the energy channeled from the $1/1$ EIC by non-linear couplings towards the thermal plasma and the $2/3$ interchange mode as well as the $3/4$ EPM modifies the plasma flows, pressure and EP density profiles between $r/a \approx 0.65 - 0.92$, that is to say, the middle-outer plasma region is destabilized as the experiment observations indicate.  
 
In short, the burst event triggered in the non linear simulation shows an increase of the $1/1$ EIC frequency to $9.35$ kHz, similar to the experimental measurement. Also, the complex mode structure near the $1/1$ rational surface could be caused by the destabilization of the $2/2$ and $3/3$ overtones. The inward propagation of the instability could be explained by the non linear destabilization of the $2/3$ and $3/4$ EPMs due to the energy channeled from the $1/1$ EIC. Therefore, the multiple resonances triggered at nearby radial locations and frequency ranges could lead to an enhancement of the EP losses, a resonance overlapping already proposed and analyzed by other authors \cite{32}.

\subsection{Phase IV: stabilization}

\begin{figure}[h!]
\centering
\includegraphics[width=0.45\textwidth]{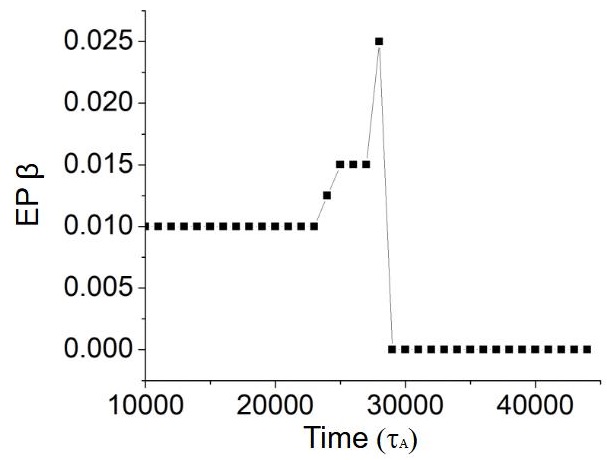}
\caption{EP $\beta$ along the simulation.}\label{FIG:12}
\end{figure}

Now, the stabilization phase (Phase IV) of the $1/1$ EIC is analyzed. For simplicity, a new simulations is performed including only the $n=1$ toroidal family in the non linear simulation, excluding the $n=2$ and $3$ toroidal mode families. The analysis target is reproducing the plasma conditions that lead to the transition from an unstable $1/1$ EIC to the RIC triggering observed in the experiment. Figure~\ref{FIG:12} shows the EP $\beta$ along the simulation. EP $\beta = 0.01$ between $t = 1.0 - 2.3 \cdot 10^{4} \tau_{A}$, increasing to $0.025$ at $t =2.8 \cdot 10^{4} \tau_{A}$. From $t = 2.9 \cdot 10^{4} \tau_{A}$ the EP $\beta$ is set to zero reproducing the EP losses caused by the $1/1$ EIC. It should be noted that in LHD discharges not all the EP population is lost after the $1/1$ EIC is triggered, only a fraction of the EP population, thus a chain of instabilities are observed in the experiment as the EP $\beta$ exceeds or decreases below the destabilization threshold of the $1/1$ EIC. To avoid the destabilization of another $1/1$ EIC along the simulation the EP $\beta$ is set to zero, so that the study can be focused in the transition from the $1/1$ EIC to the RIC. In addition, the EP $\beta=0.025$ before setting the EP $\beta$ to zero in order to maximize the destabilizing effect of the EP on the transition between the EIC phases III to IV. That way, it is easier to analyze the effect of the EP on the RIC stability before and after the transition, leading to a more didactic evaluation of the experimental data.

\begin{figure}[h!]
\centering
\includegraphics[width=0.45\textwidth]{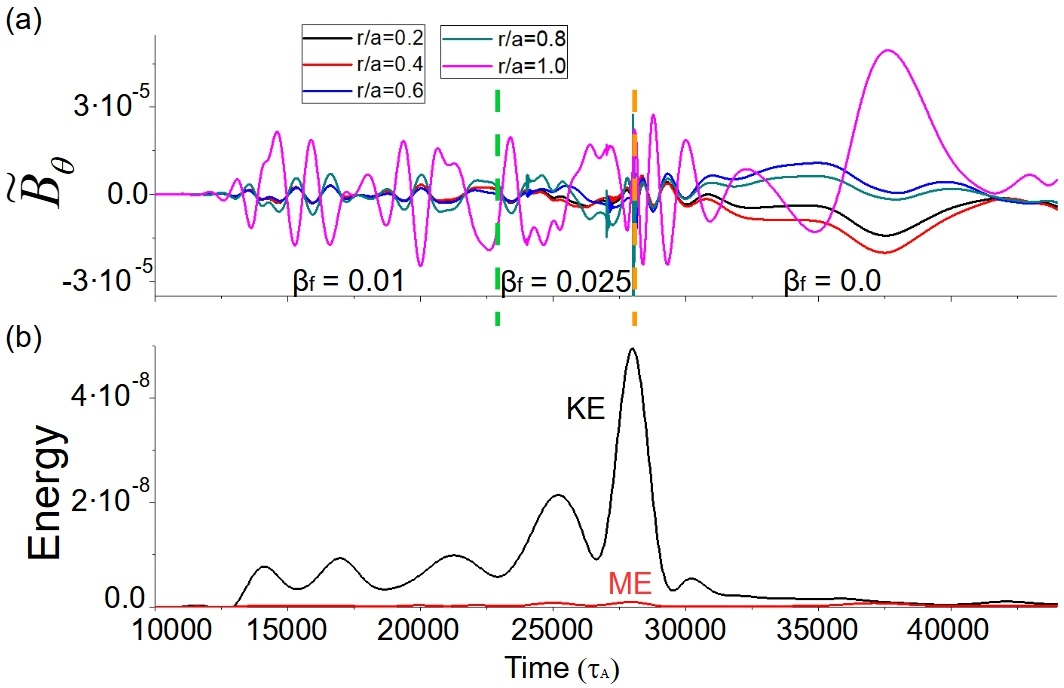}
\caption{(a) Time evolution of the poloidal component of the magnetic field perturbation at $r/a = 1.0$ (pink line), $0.8$ (cyan line),  $0.6$ (blue line), $0.4$ (red line) and $0.2$ (black line). (b) Normalized kinetic (black line) and magnetic (red) energy. The horizontal dashed green lines indicates the simulation time when the EP $\beta$ increases linearly from $0.01$ to $0.025$. The horizontal orange dashed line indicates the simulation time when the EP $\beta$ is set to zero.}\label{FIG:13}
\end{figure}

Figure~\ref{FIG:13}a shows the time evolution of $\tilde B_{\theta}$ at different radial locations in the Phase IV. From $t =1.3 \cdot 10^{4} \tau_{A}$ the plasma periphery is unstable once the EP $\beta$ threshold of the $1/1$ EIC is exceed, further enhanced as the EP $\beta$ increases. There is a transition from high frequency to low frequency $\tilde B_{\theta}$ oscillations after the EP $\beta$ is set to zero, indicating a change of the perturbation source. Figure~\ref{FIG:13}b shows an increase of the KE and ME after the $1/1$ EIC is triggered reaching its maximum once the EP $\beta = 0.025$. The KE and ME decrease after the EP $\beta$ is set to zero, almost an order of magnitude smaller with respect to the $1/1$ EIC, pointing out that the low frequency mode is potentially less hazardous for the device performance.

\begin{figure}[h!]
\centering
\includegraphics[width=0.45\textwidth]{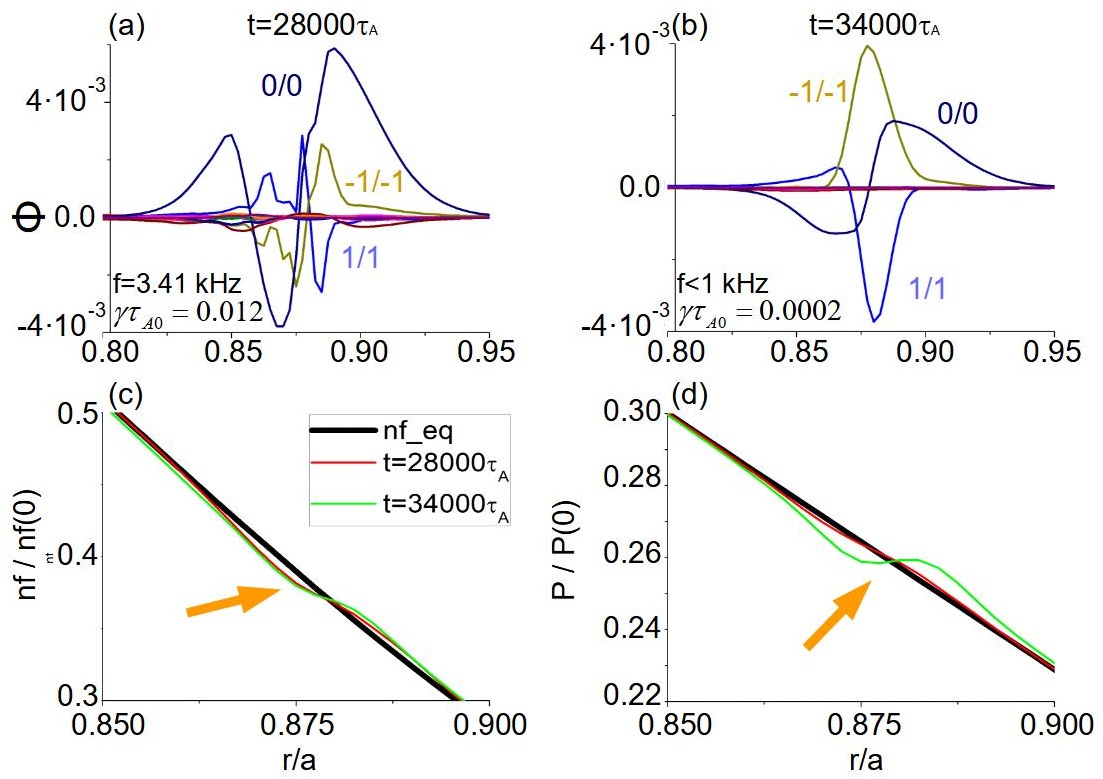}
\caption{Evolution of the electrostatic potential eigenfunction at (a) $t =2.8 \cdot 10^{4} \tau_{A}$ and (b) $t =3.4 \cdot 10^{4} \tau_{A}$. Evolution of the (c) EP density and (d) pressure profiles. The black broad line indicates the equilibrium profiles. The orange arrow indicates the radial location of the strongest deviation from the equilibrium. }\label{FIG:14}
\end{figure}

Figure~\ref{FIG:14}, panels a and b, shows the eigenfunction of the electrostatic potential during the Phase IV. At $t =2.8 \cdot 10^{4} \tau_{A}$ the $1/1$ EIC is unstable with $f = 3.4$ kHz and a growth rate of $\gamma \tau_{A0} = 0.012$. At $t =3.4 \cdot 10^{4} \tau_{A}$ the frequency of the mode decreases below $1$ kHz and the eigenfunction shows symmetric $1/1$ mode parities, thus the mode is a RIC with a growth rate almost two orders of magnitude smaller than the $1/1$ EIC. It should be noted that the amplitude of the mode $0/0$ decreases by a $60 \%$, indicating that the effect of the RIC on the equilibrium plasma flows is also smaller. The evolution of the pressure profiles shows a flattening at $t =3.4 \cdot 10^{4} \tau_{A}$, panel d, consistent with the saturation of a pressure gradient driven mode. On the other hand, the EP density profile shows only a small deviation from the equilibrium profile (panel c).
 
In summary, the simulation reproduces the $1/1$ EIC stabilization due to the decrease of the helically trapped EP population and the RIC destabilization once the stabilizing effect of the $1/1$ EIC vanishes and the pressure profile gradient builds up again at the plasma periphery.

\section{Conclusions and discussion \label{sec:conclusions}}

Two non linear simulations are performed by the FAR3d code analyzing the saturation of the $1/1$ EIC. The simulation results show a reasonable agreement with the experimental data \cite{7}. 

The non linear simulations calculate an EP $\beta$ threshold of $0.006$ for the destabilization of the $1/1$ EIC, similar to the experiment. In addition, the enhancement of the RIC and $1/1$ EIC width and amplitude as the EP $\beta$ increases is also reproduced.

\begin{figure*}[h!]
\centering
\includegraphics[width=0.9\textwidth]{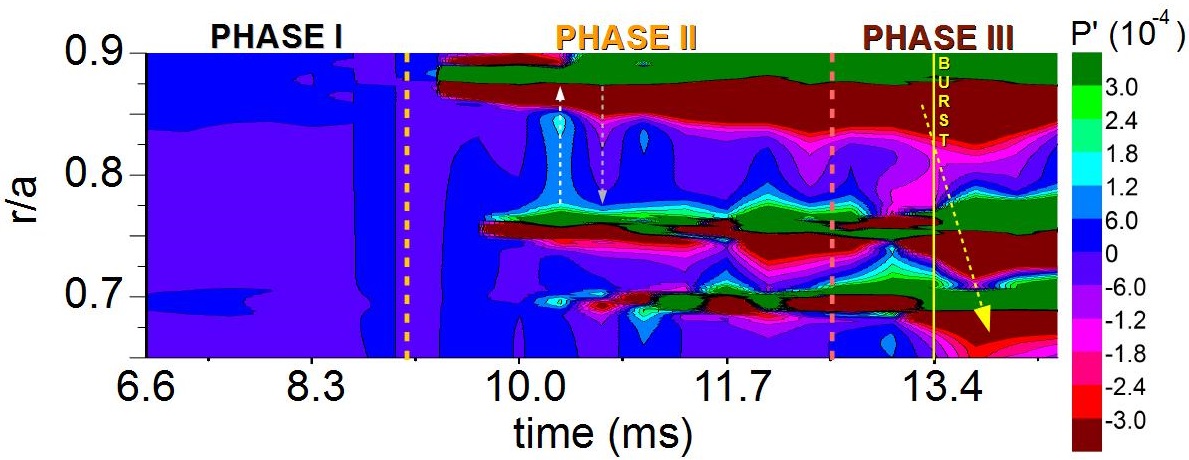}
\caption{Evolution of the pressure perturbation between $r/a = 0.65 - 0.9$. The orange dashed lines indicate the transition between the $1/1$ EIC phases. The dotted white/gray lines indicate the counter-propagating perturbations during Phase II. The solid yellow arrow indicates the burst even and the dotted yellow line the inward propagation of the $1/1$ EIC perturbation.}\label{FIG:15}
\end{figure*}

The inward propagation of the instability from the plasma periphery to the middle plasma region is observed in the simulations. The inward propagation is caused by the non linear destabilization of the modes $2/3$ and $3/4$. The $1/1$ EIC energy is channeled towards the thermal plasma, leading to the destabilization of the $2/3$ interchange mode in the phase II, as well as by non linear coupling with the $3/4$ and $2/3$ EPMs in the phases II and III, respectively. Figure~\ref{FIG:15} shows the evolution of the pressure perturbation between the middle and outer plasma region. The evolution of the pressure perturbation has similarities with the evolution of the electron temperature perturbation measured in the experiment (please see the fig. 7b of the reference \cite{7}). If the $1/1$ EIC phases of the experiment and the simulation are compared, the numerical model reproduces the enhancement of the perturbation at $r/a = 0.88$, $0.75$ and $0.68$ caused by the destabilization of the $1/1$, $3/4$ and $2/3$ modes during the Phase II. In addition, once the $3/4$ mode is destabilized, the perturbation expands outward from $r/a = 0.75$ to $0.85$ (white dotted arrow) followed by the inward expansion of the $1/1$ perturbation from $r/a = 0.9$ to $0.78$ (gray dotted arrow). Such consecutive counter-propagating perturbations are also observed in the experiment \cite{33}. After the burst event is triggered during the Phase III (solid yellow line) the $1/1$, $3/4$ and $2/3$ perturbations overlap and the instability propagates further inward (dotted yellow line), consistent with the inward propagation of the instability observed in the experiment between $r/a = 0.88$ and $0.65$. It should be noted that certain differences are observed between the experimental data and the simulation results, particularly with respect to the radial transport. The simulations show predominant horizontal contourns and some vertical mixing, although the experimental contourns of the electron temperature are mainly vertical, that is to say, the radial transport in the experiment is stronger. Present developing efforts of the numerical model are dedicated to improve the transport description reducing the discrepancy between simulation and experimental observations. Consequently, the transport analysis is out of the scope of this study. In addition, FAR3d code is a single fluid model thus electrons and ions cannot be analyzed separately, reason why the comparison between the measured electron temperature perturbation and the pressure perturbation provided by the simulation should be understood as a first order approximation. Such information is useful to analyze the inward propagation of the instability observed in the experiment, although a two fluid model is required to provide a more complete description of the burst event.

The linear simulations performed using the profiles obtained in the non linear simulation show that during the Phase II the $3/4$ perturbation is an EPM with a frequency of $12.2$ kHz and the $2/3$ perturbation an interchange mode, although the mode $2/3$ evolves to an EPM with a frequency of $5.25$ kHz during the Phase III. The $3/4$ and $2/3$ EPMs are destabilized because part of the $1/1$ EIC energy is channeled by non linear couplings toward these modes. On the other hand, the $2/3$ interchange mode is triggered because the $1/1$ EIC energy is also transferred toward the thermal plasma.

The non linear simulations also indicate the effect of the $1/1$ EIC on the equilibrium profiles, particularly by the destabilization of the $0/0$ mode. The saturation of the $1/1$ EIC begins as soon as the EP density profile flattens around the $1/1$ rational surface, leading to a decrease of the EP density profile gradient at the plasma periphery as well as the available free energy required to destabilize the mode. It should be noted that the destabilization of the $1/1$ EIC leads to a loss of the EP population and a decrease of the EP $\beta$ in the experiment, although this effect is only included in the simulation dedicated to the analysis of the stabilization phase of the $1/1$ EIC (Phase IV) for simplicity and computational efficiency. In addition, once the $1/1$ EIC is triggered the pressure profile shows a flattening around the $1/1$ rational surface and the pressure gradient at the plasma periphery is weaker. This result indicates that the stability of the EPM and pressure gradient driven modes are connected non linearly. Consequently, the RIC is stable after the $1/1$ EIC is triggered because the RIC saturates due to the flattening of the pressure profile induced via the $1/1$ EIC. On top of that, the $1/1$ EIC and the $3/4$ EPM modify the surrounding equilibrium plasma flows because the $0/0$ component of the electrostatic potential is destabilized, consistent with the experiment observations \cite{7}. On the other hand, the effect of the $1/1$ EIC on the $\rlap{-} \iota$ profile is small because there are only small changes on the radial variation of the $0/0$ component of the poloidal flow, particularly at the plasma periphery due to the proximity of the device coils (data not shown).

The simulation calculates an increase of the $1/1$ EIC frequency to $9.4$ kHz during the burst event, similar to the experimental observations. In addition, the overtones $2/2$ and $3/3$ are destabilized showing the largest growth rate of the respective toroidal families. The overtone destabilization as well as the perturbation overlapping with the $3/4$ and $2/3$ EPMs could explain the complex mode structure measured in the experiment. The inward propagation of the instability can be also tracked by the transition of the dominant mode of the $n=3$ toroidal family between the $3/3$ to the $3/4$ (mode with the largest growth rate) during the burst event, as well as the EP and pressure profiles flattening that appear around the $3/4$ and $2/3$ rational surfaces. In addition, the large decrease of the EP population measured during the burst event could be caused by the partial overlapping between the resonances of the $1/1$ EIC, $3/4$ EPM and $2/3$ EPM, leading to the enhancement of the outward transport of the helically trapped EP population \cite{32}.

The non linear simulation that analyzes the stabilization of the $1/1$ EIC caused by loss of the EP population shows the destabilization of a RIC as soon as the stabilizing effect of the $1/1$ EIC vanishes. This is because the flattening induced in the pressure profile disappears and the profile gradient recovers, exceeding the stability limit of the interchange modes at the plasma periphery. It should be noted that the energy of the RIC is almost one order of magnitude smaller with respect to the $1/1$ EIC, thus the RIC is a less dangerous instability for device performance.

Present simulations include a reduced set of toroidal families to minimize the simulation time although retaining the essential physics required to analyze the saturation phase of the $1/1$ EIC, particularly the non linear coupling between the $n=1$ and the $n=0$, $2$ and $3$ toroidal families. Nevertheless, future studies will include an extended number of toroidal families as well as the effect of the helical couplings; this is expected to have a meaningful contribution during the bursting event as a previous study indicated \cite{8}. In addition, the collective transport driven by the EIC will be evaluated by the derivation of an effective transport coefficient for the EP.

An improved LHD performance requires the attenuation or avoidance of the $1/1$ EIC. The linear stability and optimization strategies of the $1/1$ EIC are already discussed in other publications from the experimental and theoretical point of view \cite{8,25,34,35,36,37}, although a detailed study of the non linear evolution is mandatory to analyze the saturation phase of the instability. Some LHD operation scenarios require strong plasma heating by the perpendicular NBIs thus a large population of helically trapped EP is generated, exceeding the stability limit of the $1/1$ EIC. The non linear simulation results indicate that the most injurious effect of the $1/1$ EIC on the device performance takes place during the burst event. The resonance overlap caused by the non linear destabilization of the $3/4$ and $2/3$ EPMs lead to significant losses in the helically trapped EP population before thermalization, reducing the heating efficiency of the perpendicular NBIs and limiting the maximum $\beta$ reached during the discharge. Consequently, the transition from the Phase II to III of the EIC $1/1$ should be avoided. A possible indicator of the EP $\beta$ threshold for the phase II to III transition could be the electron temperature fluctuations nearby the rational surfaces $3/4$ and $2/3$, indicating a large energy channeling from the $1/1$ EIC and the destabilization of the $3/4$ and $2/3$ EPMs as precursors of the burst event. Dedicated experiments in LHD will be suggested to confirm the present study results.

\section*{Appendix}
The model equations and numerical scheme are indicated in this appendix. The model formulation assumes high aspect ratio, medium $\beta$ (of the order of the inverse aspect ratio $\varepsilon=a/R_0$), small variation of the fields and small resistivity. The plasma velocity and perturbation of the magnetic field are defined as
\begin{equation}
 \mathbf{v} = \sqrt{g} R_0 \nabla \zeta \times \nabla \Phi, \quad\quad\quad  \mathbf{B} = R_0 \nabla \zeta \times \nabla \psi,
\end{equation}
where $\zeta$ is the toroidal angle, $\Phi$ is a stream function proportional to the electrostatic potential, and $\tilde \psi$ is the perturbation of the poloidal flux.

The equations, in dimensionless form, are
\begin{equation}
\frac{\partial \tilde \psi}{\partial t} =  \sqrt{g} B \nabla_\| \Phi  + \eta \varepsilon^2 J \tilde J^\zeta - \left(\frac{1}{\rho}\frac{\partial \tilde \psi}{\partial \theta} \frac{\partial}{\partial \rho} + \frac{\partial \tilde \psi}{\partial \rho}\frac{1}{\rho}\frac{\partial}{\partial \theta} \right) \Phi
\end{equation}
\begin{eqnarray} 
\frac{{\partial \tilde U}}{{\partial t}} =  - v_{\zeta,eq} \frac{\partial \tilde U}{\partial \zeta} \nonumber\\
+ \sqrt{g} B  \nabla_{||} \tilde J^{\zeta} - \frac{1}{\rho} \left( \frac{\partial J_{eq}}{\partial \rho} \frac{\partial \tilde \psi}{\partial \theta} - \frac{\partial J_{eq}}{\partial \theta} \frac{\partial \tilde \psi}{\partial \rho}    \right) \nonumber\\
- {\frac{\beta_0}{2\varepsilon^2} \sqrt{g} \left( \nabla \sqrt{g} \times \nabla \tilde p \right)^\zeta } -  {\frac{\beta_f}{2\varepsilon^2} \sqrt{g} \left( \nabla \sqrt{g} \times \nabla \tilde n_f \right)^\zeta } \nonumber\\
- \tilde v \cdot \nabla \tilde U -  \left(\frac{1}{\rho}\frac{\partial \tilde \psi}{\partial \theta} \frac{\partial}{\partial \rho} + \frac{\partial \tilde \psi}{\partial \rho}\frac{1}{\rho}\frac{\partial}{\partial \theta} \right) \tilde J^{\zeta} + D_{U} \nabla^{2}_{\perp} \tilde U
\end{eqnarray} 
\begin{eqnarray}
\label{pressure}
\frac{\partial \tilde p}{\partial t} = - v_{\zeta,eq} \frac{\partial \tilde p}{\partial \zeta} + \frac{dp_{eq}}{d\rho}\frac{1}{\rho}\frac{\partial \tilde \Phi}{\partial \theta} \nonumber\\
 +  \Gamma p_{eq}  \left[{\left( \nabla \sqrt{g} \times \nabla \tilde \Phi \right)^\zeta - \nabla_\|  \tilde v_{\| th} }\right] \nonumber\\
+ \frac{\partial \tilde p}{\partial \rho} \frac{1}{\rho} \frac{\partial \tilde \Phi}{\partial \theta} - \frac{1}{\rho} \frac{\partial \tilde p}{\partial \theta} \frac{\partial \tilde \Phi}{\partial \rho} - \Gamma \tilde p \left( \nabla \sqrt{g} \times \nabla \tilde \Phi \right)^{\zeta} \nonumber\\ 
- \frac{\beta_{0}B}{2n_{0}(J-\rlap{-} \iota I)}\left( \frac{\partial}{\partial \theta} - \rlap{-} \iota \frac{\partial}{\partial \zeta} \right) \tilde p \tilde v_{||,th} + D_{p} \nabla^{2}_{\perp} \tilde p
\end{eqnarray} 
\begin{eqnarray}
\label{velthermal}
\frac{{\partial \tilde v_{\| th}}}{{\partial t}} = - v_{\zeta,eq} \frac{\partial \tilde v_{||th}}{\partial \zeta} -  \frac{\beta_0}{2n_{0,th}} \nabla_\| p \nonumber\\
+ \frac{\partial \tilde v_{||,th}}{\partial \rho} \frac{1}{\rho} \frac{\partial \tilde \Phi}{\partial \theta} - \frac{1}{\rho} \frac{\partial \tilde v_{||,th}}{\partial \theta} \frac{\partial \tilde \Phi}{\partial \rho} \nonumber\\
+ \frac{\beta_{0}B}{2n_{0}(J-\rlap{-} \iota I)}\left( \frac{\partial \tilde \psi}{\partial \rho} \frac{1}{\rho} \frac{\partial \tilde p}{\partial \theta} - \frac{1}{\rho} \partial {\tilde \psi}{\partial \theta} \frac{\partial \tilde p}{\partial \rho} \right) \nonumber\\ 
+ \frac{B}{(J-\rlap{-} \iota I)} \left( \frac{\partial}{\partial \theta} - \rlap{-} \iota \frac{\partial}{\partial \zeta} \right) v^{2}_{||,th} + D_{vth} \nabla^{2}_{\perp} \tilde v_{||,th}
\end{eqnarray}
\begin{eqnarray}
\label{nfast}
\frac{{\partial \tilde n_f}}{{\partial t}} = - v_{\zeta,eq} \frac{\partial \tilde n_{f}}{\partial \zeta} - \frac{v_{th,f}^2}{\varepsilon^2 \omega_{cy}}\ \Omega_d (\tilde n_f) - n_{f0} \nabla_\| \tilde v_{\| f}   \nonumber\\
-  n_{f0} \, \Omega_d (\tilde \Phi) + n_{f0} \, \Omega_* (\tilde  \Phi) \nonumber\\
+ \frac{\partial \tilde n_{f}}{\partial \rho} \frac{1}{\rho} \frac{\partial \tilde \Phi}{\partial \theta} - \frac{1}{\rho} \frac{\partial \tilde n_{f}}{\partial \theta} \frac{\partial \tilde \Phi}{\partial \rho} \nonumber\\
+ \frac{B}{(J-\rlap{-} \iota I)}\left( \frac{\partial}{\partial \theta} - \rlap{-} \iota \frac{\partial}{\partial \zeta} \right) \tilde n_{f} \tilde v_{||,f} + D_{nf} \nabla^{2}_{\perp} \tilde n_{f}
\end{eqnarray}
\begin{eqnarray}
\label{vfast}
\frac{{\partial \tilde v_{\| f}}}{{\partial t}} = - v_{\zeta,eq} \frac{\partial \tilde v_{||f}}{\partial \zeta}  -  \frac{v_{th,f}^2}{\varepsilon^2 \omega_{cy}} \, \Omega_d (\tilde v_{\| f}) \nonumber\\
- \left( \frac{\pi}{2} \right)^{1/2} v_{th,f} \left| \nabla_\| \tilde v_{\| f}  \right| - \frac{v_{th,f}^2}{n_{f0}} \nabla_\| n_f + v_{th,f}^2 \, \Omega_* (\tilde \psi) \nonumber\\
+ \frac{\partial \tilde v_{||,f}}{\partial \rho} \frac{1}{\rho} \frac{\partial \tilde \Phi}{\partial \theta} - \frac{1}{\rho} \frac{\partial \tilde v_{||,f}}{\partial \theta} \frac{\partial \tilde \Phi}{\partial \rho} \nonumber\\
+ \frac{B}{(J-\rlap{-} \iota I)}\left( \frac{\partial}{\partial \theta} - \rlap{-} \iota \frac{\partial}{\partial \zeta} \right) \tilde v^{2}_{||,f} + D_{vf} \nabla^{2}_{\perp} \tilde v_{||,f}
\end{eqnarray}
Equation (2) is derived from Ohm’s law coupled with Faraday’s law, equation (3) is obtained from the toroidal component of the momentum balance equation after applying the operator $\nabla \wedge \sqrt{g}$, equation (4) is obtained from the thermal plasma continuity equation with compressibility effects and equation (5) is obtained from the parallel component of the momentum balance, the equations (6) and (7) are moments of the kinetic equation for the energetic particles. Here, $U =  \sqrt g \left[{ \nabla  \times \left( {\rho _m \sqrt g {\bf{v}}} \right) }\right]^\zeta$ is the toroidal component of the vorticity, $\rho_m$ the ion mass density, $\rho = \sqrt{\phi_{N}}$ the effective radius with $\phi_{N}$ the normalized toroidal flux and $\theta$ the poloidal angle. The perturbation of the toroidal current density $\tilde J^{\zeta}$ is defined as:
\begin{eqnarray}
\tilde J^{\zeta} =  \frac{1}{\rho}\frac{\partial}{\partial \rho} \left(-\frac{g_{\rho\theta}}{\sqrt{g}}\frac{\partial \tilde \psi}{\partial \theta} + \rho \frac{g_{\theta\theta}}{\sqrt{g}}\frac{\partial \tilde \psi}{\partial \rho} \right) \nonumber\\
- \frac{1}{\rho} \frac{\partial}{\partial \theta} \left( \frac{g_{\rho\rho}}{\sqrt{g}}\frac{1}{\rho}\frac{\partial \tilde \psi}{\partial \theta} + \rho \frac{g_{\rho \theta}}{\sqrt{g}}\frac{\partial \tilde \psi}{\partial \rho} \right)
\end{eqnarray}
$v_{||th}$ is the parallel velocity moment of the thermal plasma and $v_{\zeta,eq}$ is the equilibrium toroidal rotation. $\beta_{0}$ is the equilibrium $\beta$ at the magnetic axis, $\beta_{f}$ is the maximum value of the EP $\beta$ (located at the magnetic axis in the on-axis cases but not in the off-axis cases) and $n_{f0}$ is the EP radial density profile normalized to the local maxima. $\Phi$ is normalized to $a^2B_{0}/\tau_{A0}$ and $\tilde\psi$ to $a^2B_{0}$ with $\tau_{A0}$ the Alfv\' en time $\tau_{A0} = R_0 (\mu_0 \rho_m)^{1/2} / B_0$. The radius $\rho$ is normalized to a minor radius $a$; the resistivity to $\eta_0$ (its value at the magnetic axis); the time to the Alfv\' en time; the magnetic field to $B_0$ (the averaged value at the magnetic axis); and the pressure to its equilibrium value at the magnetic axis. The Lundquist number $S$ parameter is the ratio of the resistive time $\tau_{R} = a^2 \mu_{0} / \eta_{0}$ to the Alfv\' en time. $\rlap{-} \iota$ is the rotational transform, $v_{th,f} = \sqrt{T_{f}/m_{f}}/v_{A0}$ is the radial profile of the energetic particle thermal velocity normalized to the Alfv\' en velocity at the magnetic axis $v_{A0}$ and $\omega_{cy}$ the energetic particle cyclotron frequency normalized to $\tau_{A0}$. $q_{f}$ is the charge, $T_{f}$ is the radial profile of the effective EP temperature and $m_{f}$ is the mass of the EP. The $\Omega$ operators are defined as:
\begin{eqnarray}
\label{eq:omedrift}
\Omega_d = \frac{\epsilon^2 \pi \rho^2 \omega_{b}}{d_{b}} \left[ \frac{\partial}{\partial \theta} \left( \frac{1}{\sqrt{g}} \right) \right]^{-1} \cdot \nonumber\\
\Bigg\{ \frac{1}{2 B^4 \sqrt{g}}  \left[  \left( \frac{I}{\rho} \frac{\partial B^2}{\partial \zeta} - J \frac{1}{\rho} \frac{\partial B^2}{\partial \theta} \right) \frac{\partial}{\partial \rho}\right] \nonumber\\
-   \frac{1}{2 B^4 \sqrt{g}} \left[ \left( \rho \beta_* \frac{\partial B^2}{\partial \zeta} - J \frac{\partial B^2}{\partial \rho} \right) \frac{1}{\rho} \frac{\partial}{\partial \theta} \right] \nonumber\\ 
+ \frac{1}{2 B^4 \sqrt{g}} \left[ \left( \rho \beta_* \frac{1}{\rho} \frac{\partial B^2}{\partial \theta} -  \frac{I}{\rho} \frac{\partial B^2}{\partial \rho} \right) \frac{\partial}{\partial \zeta} \right] \Bigg\}
\end{eqnarray}

\begin{eqnarray}
\label{eq:omestar}
\Omega_* = \frac{1}{B^2 \sqrt{g}} \frac{1}{n_{f0}} \frac{d n_{f0}}{d \rho} \left( \frac{I}{\rho} \frac{\partial}{\partial \zeta} - J \frac{1}{\rho} \frac{\partial}{\partial \theta} \right).
\end{eqnarray}
Here the $\Omega_{d}$ operator models the average drift velocity of the helically trapped particle and $\Omega_{*}$ models the diamagnetic drift frequency. The parameter $\omega_{b}=100$ kHz indicates the bounce frequency and $d_{b}=0.01$ m the bounce length of the helically trapped EP guiding center. For more details regarding the derivation of the average drift velocity operator please see Ref.\cite{8}.

Each equation has a perpendicular dissipation term with the characteristic coefficients $D_{i} = 2 \cdot 10^{-8} $ normalized to $a^{2}/\tau_{A0}$ ($i = U, p, vth, nf, vf$ is the perturbation index).

We also define the parallel gradient, perpendicular gradient squared (lowest order) and curvature operators as
\begin{equation}
\label{eq:gradpar}
\nabla_\| f = \frac{1}{B \sqrt{g}} \left( \frac{\partial \tilde f}{\partial \zeta} +  \rlap{-} \iota \frac{\partial \tilde f}{\partial \theta} - \frac{\partial f_{eq}}{\partial \rho}  \frac{1}{\rho} \frac{\partial \tilde \psi}{\partial \theta} + \frac{1}{\rho} \frac{\partial f_{eq}}{\partial \theta} \frac{\partial \tilde \psi}{\partial \rho} \right)
\end{equation}
\begin{eqnarray}
\label{eq:gradperp}
\nabla^{2}_{\perp} = \frac{1}{a^{2}}\frac{1}{\sqrt{g}} \Bigg\{ \frac{1}{\rho} \frac{\partial}{\partial \rho} \left( \rho g_{\theta \theta} \frac{\partial}{\partial \rho} - g_{\rho \theta} \frac{\partial}{\partial \theta}   \right)  \nonumber\\ 
+ \frac{1}{\rho} \frac{\partial}{\partial \theta} \left( -g_{\rho \theta} \frac{\partial}{\partial \rho} - g_{\rho \rho} \frac{1}{\rho} \frac{\partial}{\partial \theta} \right) \Bigg\} 
\end{eqnarray}
\begin{equation}
\label{eq:curv}
\sqrt{g} \left( \nabla \sqrt{g} \times \nabla \tilde f \right)^\zeta = \frac{\partial \sqrt{g} }{\partial \rho}  \frac{1}{\rho} \frac{\partial \tilde f}{\partial \theta} - \frac{1}{\rho} \frac{\partial \sqrt{g} }{\partial \theta} \frac{\partial \tilde f}{\partial \rho}
\end{equation}
with the Jacobian of the transformation,
\begin{equation}
\label{eq:Jac}
\frac{1}{\sqrt{g}} = \frac{B^2}{\varepsilon^2 (J+ \rlap{-} \iota I)}.
\end{equation}

Equations~\ref{pressure} and~\ref{velthermal} introduce the parallel momentum response of the thermal plasma. These are required for coupling to the geodesic acoustic waves, accounting for the geodesic compressibility in the frequency range of the geodesic acoustic mode (GAM) \cite{38,39}. The coupling between the equations of the EP and thermal plasma is done in the equation of the perturbation of the toroidal component of the vorticity (eq. 3), particularly through the fifth term on the right side, introducing the EP destabilizing effect caused by the gradient of the fluctuating EP density.

Equilibrium flux coordinates $(\rho, \theta, \zeta)$ are used. Here, $\rho$ is a generalized radial coordinate proportional to the square root of the toroidal flux function, and normalized to the unity at the edge. The flux coordinates used in the code are those described by Boozer \cite{40}, and $\sqrt g$ is the Jacobian of the coordinate transformation. All functions have equilibrium and perturbation components represented as: $ A = A_{eq} + \tilde{A} $.

The FAR3d code uses finite differences in the radial direction and Fourier expansions in the two angular variables. The numerical scheme is semi-implicit in the linear terms and explicit in the non-linear terms using a two semi-steps method to ensure $(\Delta t)^3$ accuracy. The finite Larmor radius and the electron-ion Landau damping effects are excluded from the simulations for simplicity.

\ack

The authors would like to thank the LHD technical staff for their contributions in the operation and maintenance of LHD. This work was supported by the Comunidad de Madrid under the project 2019-T1/AMB-13648, Comunidad de Madrid - multiannual agreement with UC3M (“Excelencia para el Profesorado Universitario” - EPUC3M14 ) - Fifth regional research plan 2016-2020 and NIFS07KLPH004 

\hfill \break

\end{document}